\documentclass[journal,twoside,web]{ieeecolor}
\usepackage{generic}
\usepackage{cite}
\usepackage{amsmath,amssymb,amsfonts}
\usepackage{algorithmic}
\usepackage{graphicx}
\usepackage{textcomp}
\def\BibTeX{{\rm B\kern-.05em{\sc i\kern-.025em b}\kern-.08em
    T\kern-.1667em\lower.7ex\hbox{E}\kern-.125emX}}
\begin{document}
\title{Self-Powered, Highly Sensitive, High Speed Photodetection Using ITO/WSe$_2$/SnSe$_2$ Vertical Heterojunction}
\author{Krishna Murali and Kausik Majumdar
\thanks{This work was supported in part by a grant under Indian Space Research Organization (ISRO), by the grants under Ramanujan Fellowship, Early Career Award, and Nano Mission from the Department of Science and Technology (DST), and by a grant from MHRD, MeitY and DST Nano Mission through NNetRA.}
\thanks{K. Murali and K. Majumdar are with the Department of Electrical Communication Engineering, Indian Institute of Science, Bangalore 560012, India (e-mail: kausikm@iisc.ac.in). }
}

\maketitle

\begin{abstract}
Two dimensional transition metal di-chalcogenides (TMDCs) are promising candidates for ultra-low intensity photodetection. However, the performance of these photodetectors is usually limited by ambience induced rapid performance degradation and long lived charge trapping induced slow response with a large persistent photocurrent when the light source is switched off. Here we demonstrate an indium tin oxide (ITO)/WSe$_2$/SnSe$_2$ based vertical double heterojunction photoconductive device where the photo-excited hole is confined in the double barrier quantum well, whereas the photo-excited electron can be transferred to either the ITO or the SnSe$_2$ layer in a controlled manner. The intrinsically short transit time of the photoelectrons in the vertical double heterojunction helps us to achieve high responsivity in excess of $1100$ A/W and fast transient response time on the order of $10$ $\mu$s. A large built-in field in the WSe$_2$ sandwich layer results in photodetection at zero external bias allowing a self-powered operation mode. The encapsulation from top and bottom protects the photo-active WSe$_2$ layer from ambience induced detrimental effects and substrate induced trapping effects helping us to achieve repeatable characteristics over many cycles.
\end{abstract}


\section{Introduction}
\label{sec:introduction}
\IEEEPARstart{W}{eak} intensity light detection in the visible and infrared range is of utmost technological importance in multiple fields encompassing remote sensing, public safety, medical instrumentation, space, military and industry applications. Transition metal di-chalcogenides (TMDCs) are layered two-dimensional materials \cite{Butler2013} which exhibit an extraordinary light absorbtion in spite of being nanometer thick. These materials show excellent gate tunability, moderate in-plane carrier mobility, and can be deposited in an inexpensive way (for example, chemical vapour deposition) on silicon based substrates. Further, unlike bulk semiconductors, different layered materials can be seamlessly integrated in a vertical heterojunction stack without having to worry about the lattice mismatch between two materials \cite{Novoselov2012, Geim2013, Li2016}. Consequently, TMDCs (such as MoS$_2$, MoSe$_2$, WS$_2$ and WSe$_2$) have been extensively explored in the recent past as promising candidates to achieve low cost, sensitive photodetectors \cite{Koppens2014, Choi2014, Groenendijk2014, Esmaeili-Rad2013, Su2015a, Zheng2016, Dhanabalan2016, Xia2014a, Ghosh2018a, Kallatt2018}.

Although photodetection with large responsivity has been demonstrated using TMDCs, there are two important bottlenecks that need to be overcome for these photodetection schemes to become technologically viable. First, the response time of most of the high gain photodetectors is very large owing to long lived traps \cite{Jacobs-Gedrim2014a, Kallatt2016, Krishna2018}. These traps originate both from the active medium as well as the substrate supporting the active material \cite{Krishna2018}. When one type of photogenerated carriers (either electron or hole) is captured by the traps in the active material, to maintain charge neutrality of the system, the other type of carrier experiences multiple reinjection until the trapped carrier is eliminated from the active medium either by contact or by recombination. In addition, the trapped carriers in the TMDC film or in the substrate can also lead to a photo-gating effect, and a photocurrent keeps flowing until the trapped carrier is released \cite{Kallatt2016, Furchi2014}. Both these effects result in a large gain, however, associated with a slow transient response of the photocurrent in response to light.

The second problem arises from the ultra-thin nature of the TMDC films, providing a large surface to volume ratio. This results in exposure of the active medium to surroundings, leading to ambience (including moisture effect) induced detrimental effects and hence poorly repeatable characteristics.

In order to overcome these bottlenecks, here we demonstrate a vertical heterojunction of WSe$_2$/SnSe$_2$ capped with indium tin oxide (ITO) transparent electrode. The transparent ITO layer allows light to pass through it to generate photo-excited carriers in the buried WSe$_2$ layer, while protecting the heterojunction from changing ambient conditions, allowing repeatable characteristics over many cycles. On the other hand, SnSe$_2$ eliminates the substrate induced trapping effects by isolating the WSe$_2$ layer from the SiO$_2$/Si substrate. Both SnSe$_2$ \cite{Krishna2018, Murali2018} and ITO being highly conducting material, act as closely spaced photocarrier collection medium. Such small separation of collection medium, as defined by the thickness of WSe$_2$, is difficult to achieve in lithography limited planar structures, and reduces the transit time of the photogenerated carriers leading to fast response and high gain simultaneously. In addition, asymmetric band offsets at the SnSe$_2$/WSe$_2$ and ITO/WSe$_2$ interfaces induce a strong built-in field, which enables high responsivity at zero external bias. We report a large zero-bias responsivity of $16.45$ A/W, which increases to $1139$ A/W at 0.4 V. The devices exhibit repeatable performance over many cycles with negligible persistent photocurrent, sub-$10$ $\mu$s rise time, and $10$-$30$ $\mu$s fall time.

The rest of the paper is organized as follows: We discuss the detailed features of the vertical design in section \ref{sec:design}. Followed by this, we explain the photoelectron transfer mechanism in the heterojunction using photoluminescence and Raman experiments in section \ref{sec:charge_transfer}. The fabrication and electrical characterization details of the devices are explained in section \ref{sec:fab} and the photodetection performance is discussed in section \ref{sec:performance}. We conclude the paper in section \ref{sec:conclusion}.
\section{Device Design: Why Vertical Heterojunction?}\label{sec:design}
\begin{figure}[!hbt]
\centering
\includegraphics[scale=0.25] {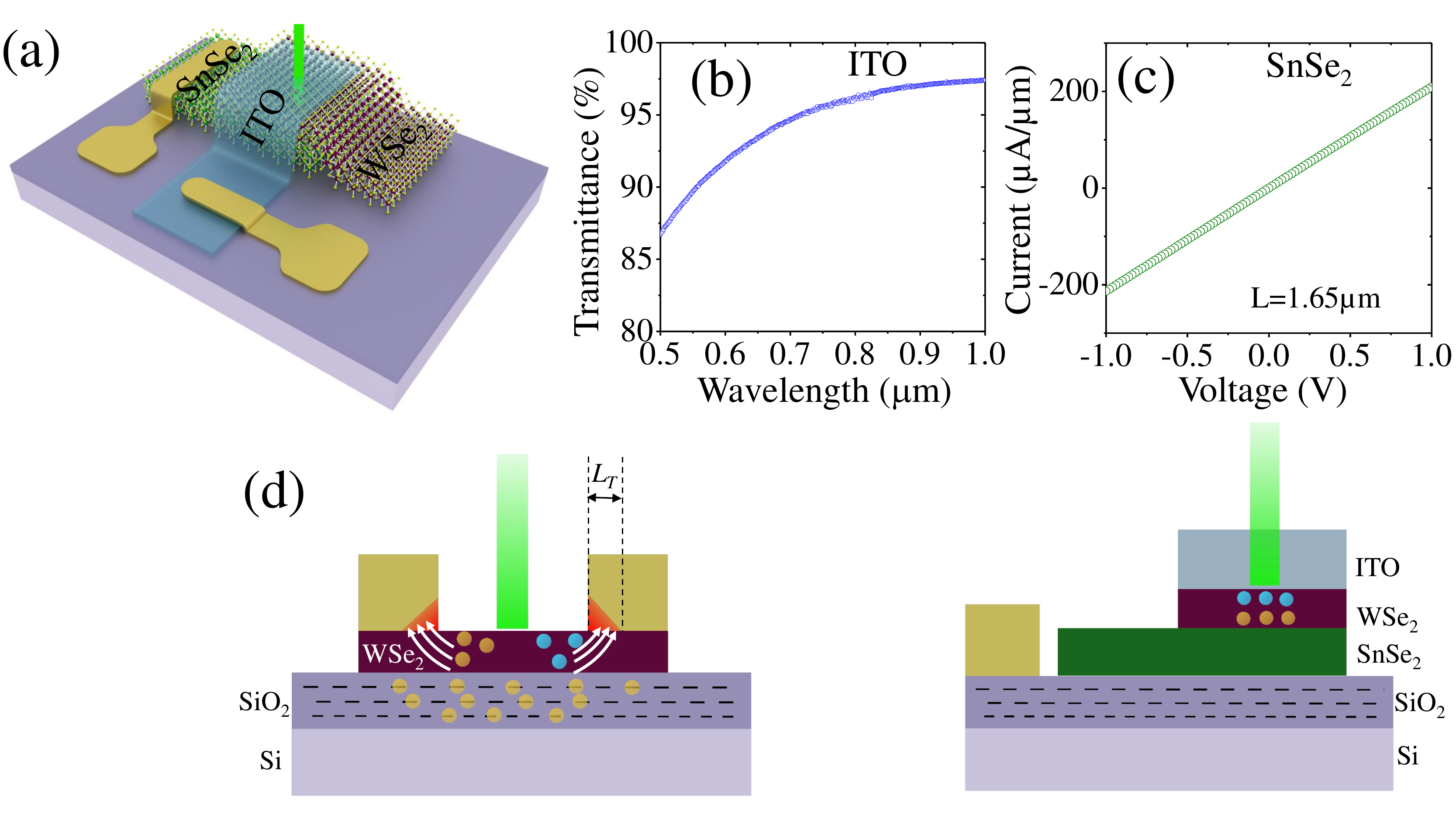}
\caption{(a) Schematic diagram of the proposed vertical device. (b) Wavelength dependent transmittance of ITO film. (c) Current-voltage characteristics of SnSe$_2$ showing high conductivity. (d) Left panel: photo-excited carrier separation and trapping in SiO$_2$ substrate in a planar structure. The dashed lines represent trap states in SiO$_2$. Current crowding at the contact interface is represented by the red region. $L_T$ is the transfer length of the contact interface. Right panel: The vertical structure with negligible current crowding and no SiO$_2$ induced trapping effect.}\label{fig:schematic}
\end{figure}

Fig. \ref{fig:schematic}(a) shows a schematic diagram of the device where a vertical heterojunction is created between SnSe$_2$ and WSe$_2$, with a transparent conducting layer ITO on top. The wavelength dependent transmittance of ITO layer is shown in Fig. \ref{fig:schematic}(b).
Highly conducting ITO and SnSe$_2$ films act as carrier collection layers. The current-voltage characteristics of a $30$ nm thick SnSe$_2$ film is shown in Fig. \ref{fig:schematic}(c). The resistivity of the ITO film measured is $2.7\times 10^{-5}$ $\Omega$m. This vertical structure provides certain unique advantages as described below. The situations are schematically illustrated in the left and right panels of Fig. \ref{fig:schematic}(d).
\begin{enumerate}
  \item \emph{Short transit time:} The WSe$_2$ active layer is sandwiched by closely separated conducting layers, where the separation is defined by the thickness of WSe$_2$. Such a small separation is difficult to achieve using lithography in a planar structure, and allows to achieve very short transit time of electrons ($\tau_{tr}$). On the other hand, the photo-excited holes are confined in the WSe$_2$ layer either due to valence band confinement in the quantum well produced by the double heterojunction or due to the band-tail state induced traps inside the bandgap. The photoconductive gain, defined as $G=\frac{\tau}{\tau_{tr}}$, where $\tau$ is the time a hole takes on an average to either recombine with an electron or flee from the quantum well. A reduced $\tau_{tr}$ helps to achieve a high gain while maintaining a fast response time.
  \item \emph{Repeatable characteristics:} The ITO layer on top completely seals the active WSe$_2$ layer from ambience which helps to achieve stable characteristics over many cycles without any performance deterioration. On the other hand, SnSe$_2$ screens all the defect states of SiO$_2$ substrate, thus eliminating the slow response tail and any persistent photocurrent.
  \item \emph{Large built-in field:} The asymmetric band offsets at the two interfaces of the ITO/WSe$_2$/SnSe$_2$ heterojunction creates a large built-in potential across WSe$_2$. This, coupled with the small thickness of WSe$_2$ results in a large built-in field, supporting efficient transport of photo-excited electrons. This, on one hand, allows high responsivity at zero external bias operation, and on the other hand provides a large open circuit voltage.
  \item \emph{Improved carrier collection:} It is important to suppress the series resistance to enhance the performance of the photodetector. The planar structure, as schematically illustrated in the left panel of Fig. \ref{fig:schematic}(d), allows carrier collection only in the effective contact area of $w\times L_T$ where $w$ is the contact width and $L_T$ is the transfer length \cite{Somvanshi2017a}. This results in an additional series resistance degrading performance. However, in the current vertical structure, such current crowding induced resistance is suppressed as the bending of current happens at the highly conducting SnSe$_2$ layer. This allows carrier collection over the entire photodetector area \cite{Kallatt2018}.
\end{enumerate}
\section{Charge transfer in ITO/WSe$_2$/SnSe$_2$ heterojunction}\label{sec:charge_transfer}
\begin{figure*}[!hbt]
\centering
\includegraphics[scale=0.4] {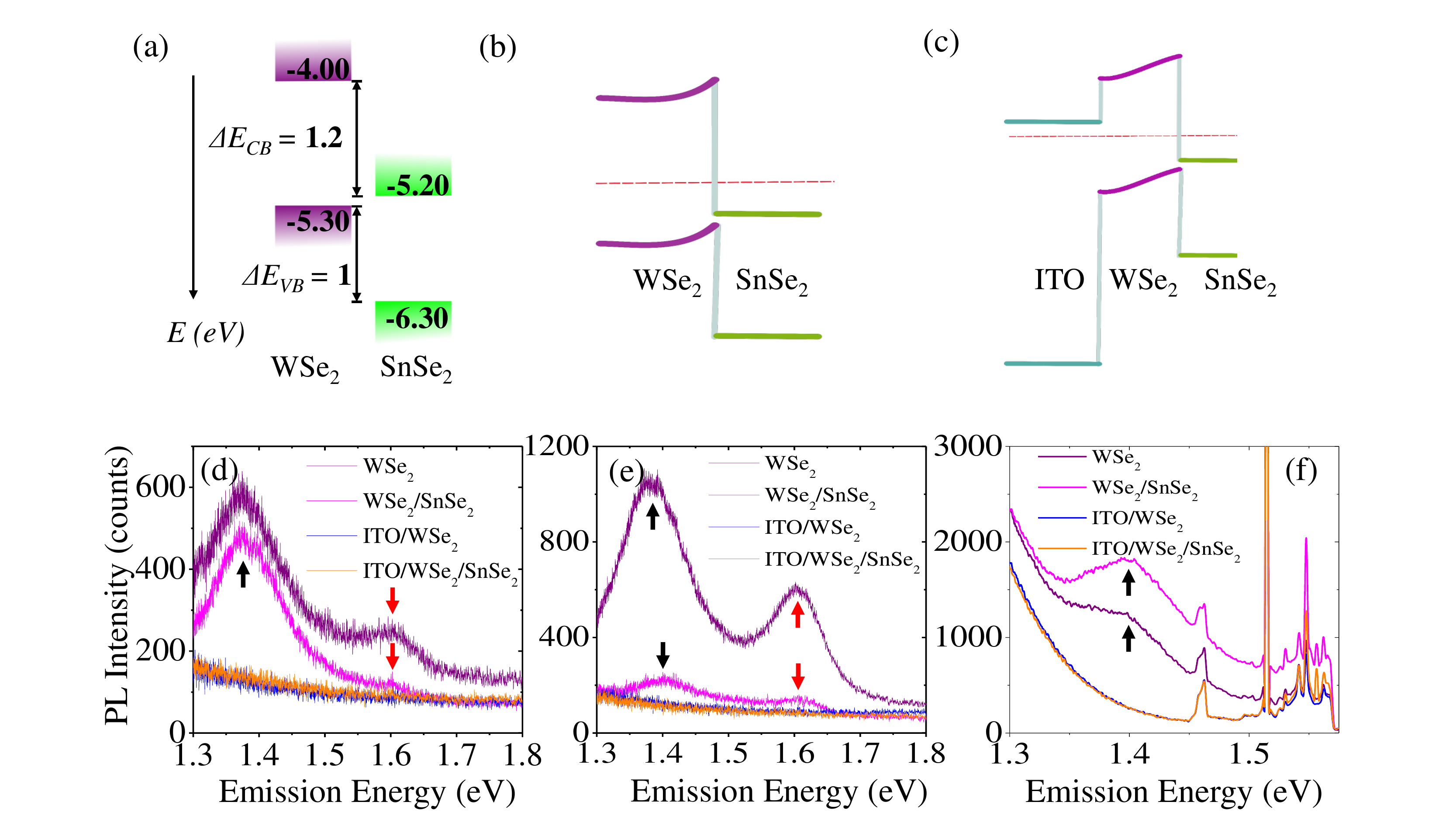}
\caption{(a) Band offset between multi-layer WSe$_2$ and SnSe$_2$. (b) Band diagram of WSe$_2$/SnSe$_2$ heterojunction with no contact on top. Degenerate doping forces negligible band bending in SnSe$_2$. (c) Band diagram of ITO/WSe$_2$/SnSe$_2$ double heterojunction showing built-in electric field in WSe$_2$, and confinement of holes in the valence band quantum well. (d)-(e) Photoluminescence spectra of the different portions of the heterojunction using $532$ nm excitation with (d) $>30$ nm WSe$_2$ and (e) $\approx 10$ nm WSe$_2$. Red and black arrows indicate direct and indirect peaks, respectively. (f) Photoluminescence spectra of the different portions of the heterojunction using $785$ nm excitation with $\approx 10$ nm WSe$_2$. The sharp peaks close to the excitation energy indicate Raman peaks from Si, SnSe$_2$ and WSe$_2$.}\label{fig:transfer}
\end{figure*}
The band offsets of WSe$_2$ and SnSe$_2$ are illustrated in Fig. \ref{fig:transfer}(a) \cite{Krishna2018, Murali2018, Schlaf1999, Lang1994, Aretouli2016}. Such a large band offset, coupled with large n-type doping of SnSe$_2$ gives rise to the WSe$_2$/SnSe$_2$ heterojunction band diagram in equilibrium as schematically shown in Fig. \ref{fig:transfer}(b). With an ITO layer on top, the band diagram is shown in Fig. \ref{fig:transfer}(c). The steep conduction band offsets at the WSe$_2$/SnSe$_2$ and WSe$_2$/ITO interfaces facilitate electron transfer from WSe$_2$ to both SnSe$_2$ and ITO, respectively, whereas the holes remain confined in the WSe$_2$ quantum well. In addition, the built-in electric field inside the WSe$_2$ layer plays an important role in the electron transfer mechanism. To get insights, we perform systematic photoluminescence (PL) experiments. First we fabricate WSe$_2$/SnSe$_2$ heterojunctions and record energy resolved PL emission from WSe$_2$, both on isolated WSe$_2$ as well as on the junction area. Next, a blanket ITO layer is coated on the sample using RF sputtering, and the PL is remeasured to see the effect of ITO. The amount of charge transfer is qualitatively inferred from the degree of quenching of the photoluminescence peaks with respect to the isolated WSe$_2$ flake.

Fig. \ref{fig:transfer}(d) depicts the acquired PL spectra of $> 30$ nm thick WSe$_2$ flake for the four different stacking variations using $532$ nm laser. The peak around $1.6$ eV, indicated by the red arrow, corresponds to the direct bandgap at the $K$ point of the Brillouin zone. The peak around $1.37$ eV, shown by the black arrow, is the indirect gap. We observe that, although there is a small suppression of direct peak at the junction, the indirect peak of WSe$_2$ remains strong in both isolated WSe$_2$ and WSe$_2$/SnSe$_2$ junction. This suggests less degree of charge transfer between the two layers and is due to a larger fraction of the thick WSe$_2$ flake residing away from the junction (top region) contribute to the PL, where charge transfer to SnSe$_2$ is not efficient, as expected from the band diagram in Fig. \ref{fig:transfer}(b). However, when ITO is deposited on top, both cases show strong quenching of PL due to an almost complete transfer of electrons from WSe$_2$ to ITO, as favoured by the built-in field in WSe$_2$ and the band offset at the ITO/WSe$_2$ interface. However, the situation changes when we use $\approx 10$ nm thick flake. Here, almost the entire WSe$_2$ region is close to the SnSe$_2$ layer, and hence the hot photoelectrons generated by $532$ nm ($2.33$ eV) photons experience a high probability of being transferred to SnSe$_2$, quenching the PL, as illustrated in Fig. \ref{fig:transfer}(e). As expected, with ITO on top, PL intensity is again completely quenched due to efficient electron transfer to ITO.

Fig. \ref{fig:transfer}(f) explains the results when the experiment is performed on the $\approx 10$ nm thick WSe$_2$ flake using $785$ nm laser excitation. Here we observe only the indirect peak in isolated WSe$_2$, as excitation is almost resonant with the direct peak energy. Note that, unlike Fig. \ref{fig:transfer}(e), we do not observe any quenching of the indirect peak in the WSe$_2$/SnSe$_2$ junction. This is because the excitation energy being much less ($1.58$ eV), we do not generate high energy photoelectrons. The photo-carriers, with energy close to the $K$ point band edge thus are not so easily transferred to SnSe$_2$. Rather, these carriers are preferably transferred to the lower energy indirect valley, hence a strong PL is maintained in the junction. As earlier, with ITO on top, the electrons are easily transferred to ITO and the PL is completely quenched.
Note that the strong Raman peaks of WSe$_2$ observed in the ITO/WSe$_2$ and ITO/WSe$_2$/SnSe$_2$ structures in Fig. \ref{fig:transfer}(f) suggest that the quality of the WSe$_2$ remains intact after ITO deposition on top.

Thus, we are able to tune the degree and direction of the photoelectron transfer from the WSe$_2$ layer by controlling the excitation wavelength and the thickness of the WSe$_2$ flake. Such controlled and efficient carrier transfer from the WSe$_2$ even without the application of any external bias forms the basis of the operation principle of the proposed photodetector, as explained in the next sections.
\section{Device Fabrication and Characterization}\label{sec:fab}
\begin{figure}[!hbt]
\centering
\includegraphics[scale=0.17] {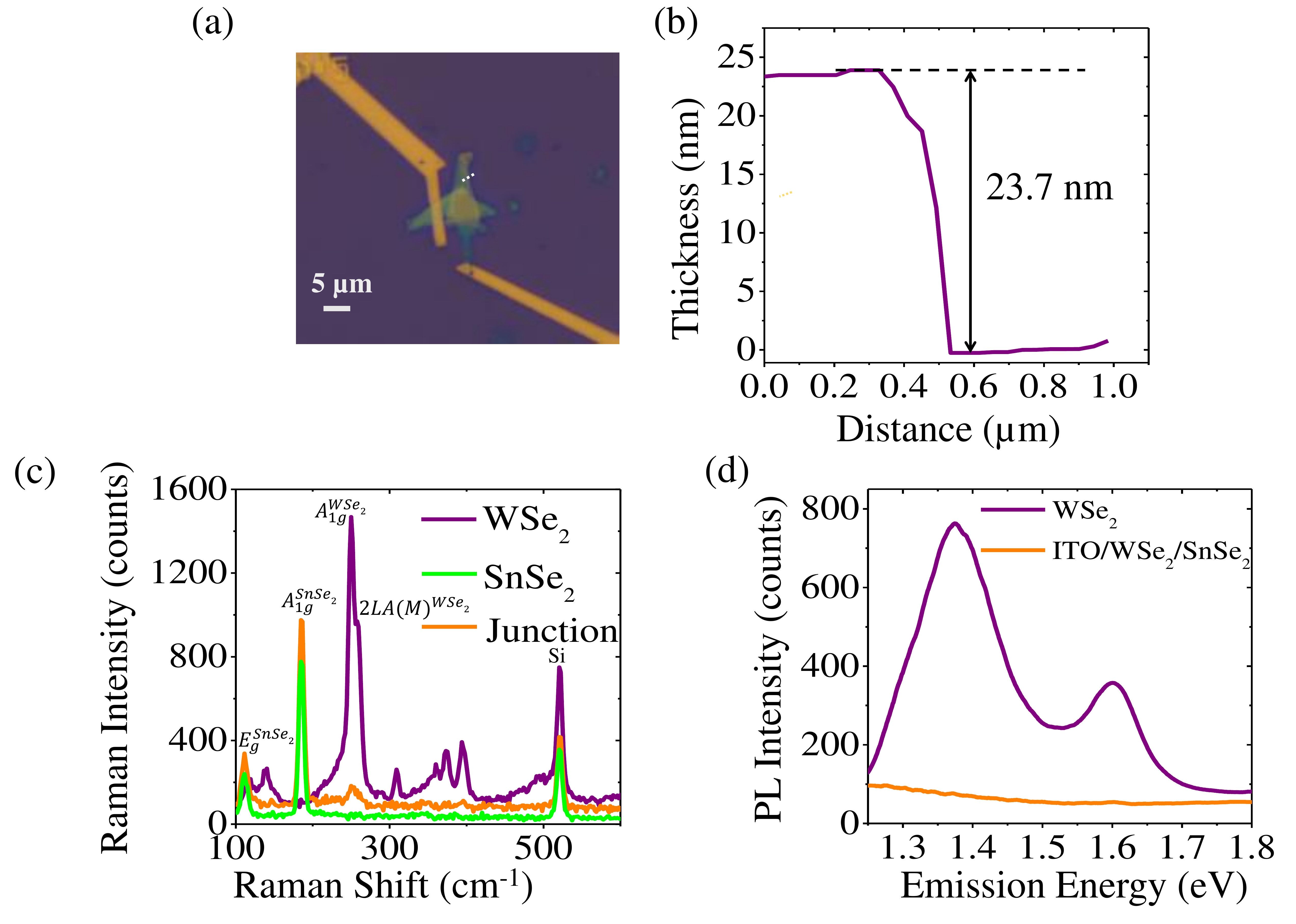}
\caption{(a) Optical image of the fabricated device ($D$1). Scale bar is $5$ $\mu$m. (b) Thickness of the WSe$_2$ film in device $D$1 obtained from AFM scanned along the white dashed line in (a). (c) Raman shift of  WSe$_2$ (purple) and SnSe$_2$ (green) in individual isolated portions. The Raman signal from the ITO/WSe$_2$/SnSe$_2$ heterojunction is shown in orange, showing strong suppression of WSe$_2$ Raman peaks. (d) PL spectra from isolated WSe$_2$ and from the junction portion using $532$ nm laser, showing strong suppression of PL intensity in the junction.}\label{fig:char}
\end{figure}
The photoconductive vertical detector is fabricated on Si substrate covered with $300$ nm SiO$_2$. The SnSe$_2$ multi-layer flake is first mechanically exfoliated on the substrate. After identifying the flake of interest, a few-layer thick WSe$_2$ flake is precisely transferred on top of the SnSe$_2$ flake using PDMS film with the help of a micromanipulator under an optical microscope. For better adhesion, we heat this stack at $180^\circ$ C for $3$ minutes in hotplate. A transparent metal electrode (ITO) is defined on top of the heterojunction using electron beam lithography, followed by deposition of $30$ nm ITO using RF sputtering and subsequent lift off. Finally, the electrodes for contacts are defined by a second electron beam lithography, followed by the deposition of Ni ($10$ nm)/Au ($50$ nm) electrodes by DC sputtering. After this, the substrate is pasted on a PCB. External wires are taken for photocurrent measurement by wire bonding these devices using Al wire from the Ni/Au contacts. Fig. \ref{fig:char}(a) shows the optical image of a representative device ($D$1) after completion of fabrication. The thickness of the WSe$_2$ film is found to be $23.7$ nm with Atomic Force Microscopy [Fig. \ref{fig:char}(b)]. Raman spectra of the isolated WSe$_2$, SnSe$_2$ and ITO/WSe$_2$/SnSe$_2$ junction are presented in Fig. \ref{fig:char}(c) using the excitation of 532 nm laser. The Raman peaks corresponding to the E$_g$ and A$_{1g}$ vibrational modes of SnSe$_2$ are observed in the isolated part of the SnSe$_2$ flake at $110$ and $185$ cm$^{-1}$, respectively. On the other hand, Raman peaks of isolated WSe$_2$ flake corresponding to A$_{1g}$ and $2$LA (M) modes are observed at $251$ and $258$ cm$^{-1}$, respectively. However, we observe that these modes are heavily suppressed in the junction area, while the SnSe$_2$ Raman peaks remain strong. The photo-excited high energy hot carriers, instead of scattering with the zone center phonon, are transferred quickly to the SnSe$_2$ layer due to built-in field, leading to a suppression of the Raman scattered light in the junction. Interestingly, this is in contrast with the strong Raman signal from the junction when $785$ nm excitation is used in Fig. \ref{fig:transfer}(f). This is possibly due to the fact that in the latter case, $785$ nm laser coherently excites the $K$ and $K^\prime$ excitons, which do not experience the built-in field due to charge neutrality, and hence not so easily transferred to SnSe$_2$. Rather these excitons scatter with WSe$_2$ phonons, giving rise to strong Raman signal. As expected from the discussion in the previous section, we observe strong quenching of WSe$_2$ PL intensity in the junction area of the device, indicating efficient photo-excited electron transfer [Fig. \ref{fig:char}(d)].

\begin{figure}[!hbt]
\centering
\includegraphics[scale=0.27] {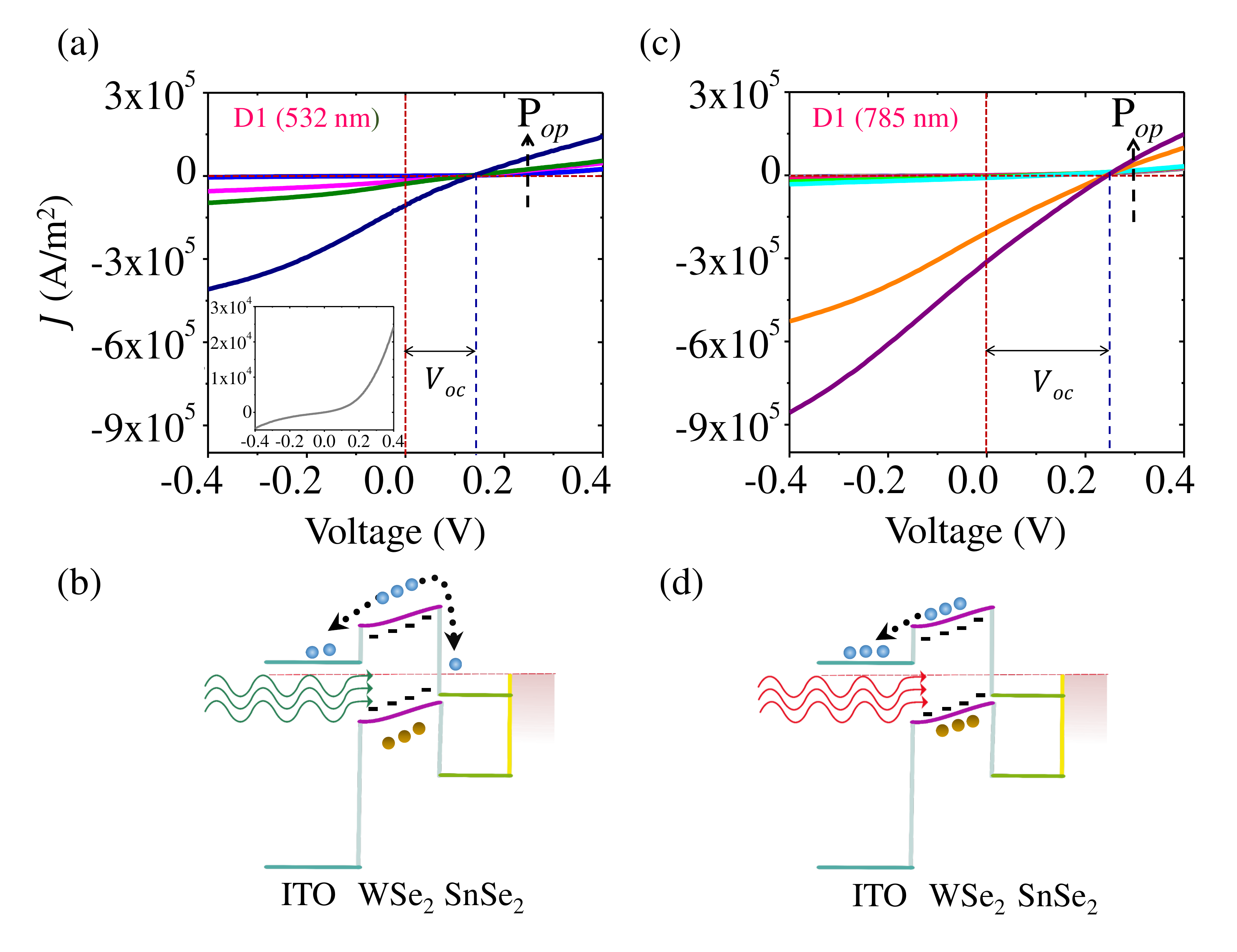}
\caption{(a), (c) Current density - Voltage characteristics of device $D$1 (area = $8.36$ $\mu$m$^2$) under illumination of (a) $532$ nm and (c) $785$ nm wavelength lasers with increasing optical power density. The Power density varies from $1.89\times10^4$ to $1.89\times10^8$ W/m$^2$ for $532$ nm and from $2.67$ to $5.35\times10^6$ W/m$^2$ for $785$ nm. Large zero bias photocurrent ($I_{sc}$) and open circuit voltage ($V_{oc}$) are observed in both cases. Inset of (a): Dark current density with applied bias. (b), (d) Schematic energy band diagram for ITO/WSe$_2$/SnSe$_2$ heterojunction for $D$1 at zero bias, with electron (blue spheres) transfer mechanism for (b) $532$ nm and (d) $785$ nm illumination. The holes (orange spheres) are confined in WSe$_2$. $532$ nm photons create higher energy photo-excited carriers than $785$ nm.}\label{fig:elec1}
\end{figure}
All photoresponse measurements are carried out using continuous wave laser excitation with wavelength of $532$ nm and $785$ nm. Electrical transport measurements are conducted on several devices with varying thickness of WSe$_2$ flakes. In this paper, all the measurements are taken in ambient air condition and applying bias on the SnSe$_2$ contact while keeping ITO side grounded. Fig. \ref{fig:elec1}(a) represents the current-voltage (I-V) characteristics of the device $D$1 under dark condition, as well as under  illumination by $532$ nm laser with varying light intensity. The corresponding band diagram at zero external bias is shown in Fig. \ref{fig:elec1}(b). The external bias dependence of the dark current density is shown separately in the inset of Fig. \ref{fig:elec1}(a). We observe from the band diagram that both under positive and negative bias conditions, the electrons need to overcome certain thermal barrier. For reverse bias (negative bias on SnSe$_2$ side), the thermal barrier for the electrons is particularly high from SnSe$_2$ to WSe$_2$. The lack of large rectification ratio (only $\approx 5$) in the device indicates that trap assisted tunneling through the WSe$_2$ film plays an important role in determining the dark current of the device.

The device shows strong photoresponse, as indicated in Fig. \ref{fig:elec1}(a). One of the striking features of the photoresponse characteristics is the large photocurrent at zero external bias, leading to an open circuit voltage $V_{oc}=0.14$ V. This is in agreement with a large built-in field existing in WSe$_2$ owing to the asymmetric band offsets at the ITO/WSe$_2$ and WSe$_2$/SnSe$_2$ interfaces, as illustrated in the band diagram shown in Fig. \ref{fig:elec1}(b). The direction of the zero bias short circuit current ($I_{sc}$) suggests that the photogenerated electrons in WSe$_2$ are transferred to the ITO layer. The photo-excited holes are confined in the WSe$_2$ quantum well by the large valence band barriers offered by both the ITO and SnSe$_2$ layers. In addition, the holes can also get trapped by the shallow band-tail states in the bandgap of WSe$_2$. These effects result in a successive reinjection of the electrons until the photogenerated hole either tunnels out of the WSe$_2$ quantum well or recombines with electrons, providing a large internal gain.

Note that the photogenerated electrons in WSe$_2$ experiences a sharp band offset on both ITO and SnSe$_2$ sides [Fig. \ref{fig:elec1}(b)]. Hence, by tuning the field inside WSe$_2$ by an external bias around $V_{oc}$, the net electron collection medium can be changed from ITO to SnSe$_2$. The device thus exhibits a strong photoresponse to both polarities of external bias.
\begin{figure}[!hbt]
\centering
\includegraphics[scale=0.27] {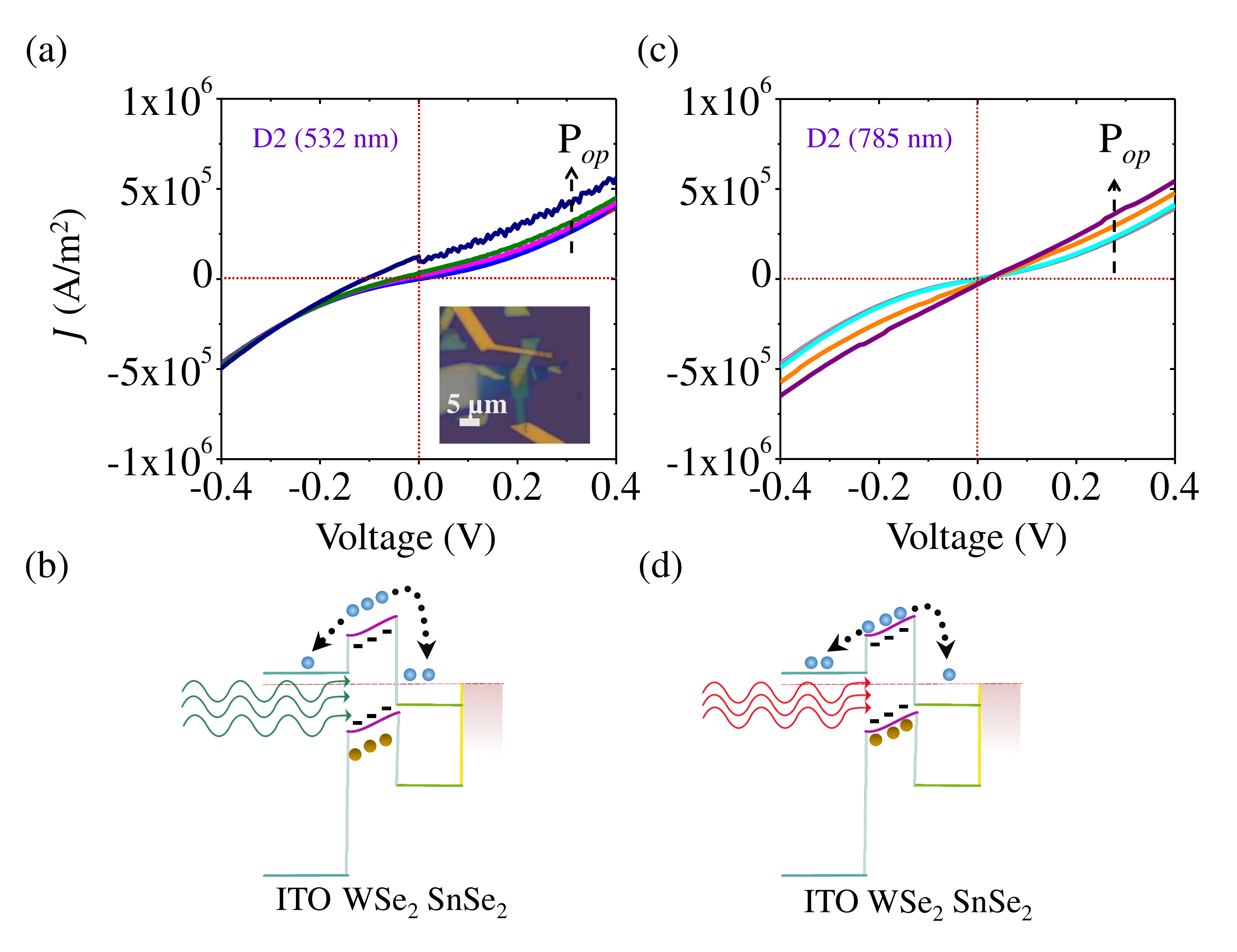}
\caption{(a), (c) Current density - Voltage characteristics of device $D$2 (area = $10.9$ $\mu$m$^2$) under illumination of (a) $532$ nm and (c) $785$ nm wavelength lasers with increasing optical power density. The Power density varies from $1.89\times10^4$ to $1.89\times10^8$ W/m$^2$ for $532$ nm and from $26.7$ to $5.35\times10^6$ W/m$^2$ for $785$ nm. The zero bias photocurrent reverses sign with changing photon wavelength. (b), (d) Schematic energy band diagram for ITO/WSe$_2$/SnSe$_2$ heterojunction for $D$2 at zero bias, with electron (blue spheres) transfer mechanism for (b) $532$ nm and (d) $785$ nm illumination. }\label{fig:elec2}
\end{figure}

Note that the device exhibits a stronger photoresponse with near-infrared $785$ nm excitation [Fig. \ref{fig:elec1}(c)-(d)]. In addition, the zero-bias response is also enhanced with an increased $V_{oc}$ of $0.25$ V. As explained earlier, photons with wavelength of $532$ nm ($E_{ph}=2.33$ eV) being of much higher energy than $785$ nm ($E_{ph}=1.58$ eV) generates higher energy hot electrons. The hot electrons generated close to the WSe$_2$/SnSe$_2$ interface thus have a high probability of falling into SnSe$_2$ for $532$ nm excitation. This results in a bidirectional segregation of the photoelectrons, suppressing the net photocurrent, as schematically explained in Fig. \ref{fig:elec1}(b). On the other hand, $785$ nm excitation generates electron-hole pair closer to the band edges [Fig. \ref{fig:elec1}(d)], and hence under zero external bias, a larger fraction of the photogenerated electrons are pulled towards the ITO layer, improving the zero bias photocurrent. Note that such a strong photocurrent at zero external bias helps us to achieve completely self-powered photodetector - an important requirement for future sensing devices that target energy saving and weight reduction. The large $V_{oc}$ also opens up the possibility for using the device in solar energy harvesting applications.

To further support the above mentioned mechanisms, we fabricate another set of devices ($D$2), which are identical to $D$1, however, but with a thinner WSe$_2$ ($\approx 10$ nm), as schematically shown in the inset of Fig. \ref{fig:elec2}(a). As expected, the dark current is enhanced in $D$2, with negligible rectification, owing to enhanced tunneling current. An interesting feature observed in the photocurrent characteristics [\ref{fig:elec2}(a)-(d)] is the change in the sign of $I_{sc}$ under zero external bias when the excitation wavelength is changed from $785$ nm to $532$ nm. Contrary to $D$1, due to thin WSe$_2$, the photogenerated electrons in WSe$_2$ are created in the close vicinity of WSe$_2$/SnSe$_2$ interface. As discussed in the previous section, compared to $785$ nm, for $532$ nm excitation, a large fraction of the high energy hot electrons are easily transferred to SnSe$_2$. In addition, a fraction of the photons completely penetrate through WSe$_2$ and get absorbed in SnSe$_2$ as well \cite{Krishna2018}. These result in a net flow of electrons towards SnSe$_2$ for $532$ nm under zero bias, leading to positive photocurrent under zero external bias. However, the net flow remains towards ITO for the $785$ nm laser due to excitation close to the band edge resulting in negative $I_{sc}$.
\section{Photodetection Performance}\label{sec:performance}
\begin{figure}[!hbt]
\centering
\includegraphics[scale=0.21] {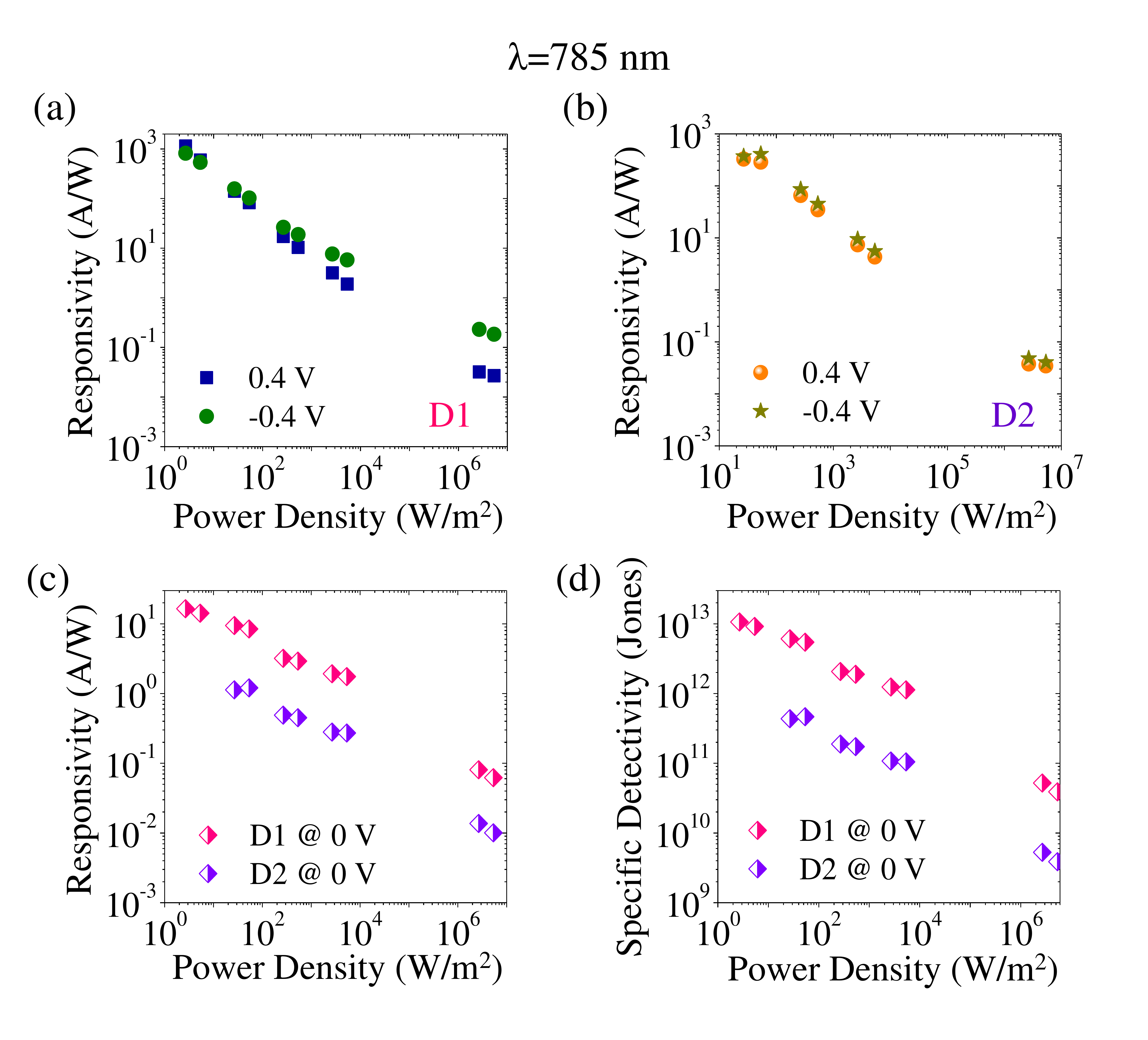}
\caption{(a),(b) Responsivity of the  devices (a) $D$1 and (b) $D$2 under illumination of 785 nm wavelength with varying power density under $\pm0.4$ V external bias. (c) Responsivity of $D$1 and $D$2 under zero external bias as a function of incident optical power density at $785$ nm.
(d) Power density dependent zero bias specific detectivity ($D^*$) of devices $D$1 and $D$2 at $\lambda=785$ nm. The area used to calculate the power density is $7.2$ $\mu$m$^2$.
}\label{fig:perf}
\end{figure}

We now discuss the performance of the devices $D$1 and $D$2 as near infrared detectors at $785$ nm wavelength. Fig. \ref{fig:perf}(a)-(b) depict the measured Responsivity ($R$) of the devices as a function of input optical power density under both positive and negative external bias ($V_{ext}$). Here responsivity is defined as $R=\frac{I_{ph}}{P_{op}}=\frac{I-I_{dark}}{P_{op}}$ where $I$ and $I_{ph}$ are the total current under illumination and the photocurrent, respectively. $I_{dark}$ is the device dark current, and $P_{op}$ is the optical power incident on the device. For $D$1, negative $V_{ext}$ produces larger $R$, particularly at higher optical power density. However, the difference is negligible for $D$2. We have obtained a responsivity of $1139$ A/W for $D$1 at $P_{op}=2.67$ W/m$^2$ with $0.4$ V external bias. For $D$2, the attained responsivity number reads $407$ A/W at $P_{op}=26.7$ W/m$^2$ and $|V_{ext}|=0.4$ V. These numbers are very impressive compared with commercially available photodetectors and indicate a large internal photoconductive gain. By noting that
\begin{equation}\label{eq:G}
R=G\eta\frac{\lambda(nm)}{1243},
\end{equation}
where $G$ and $\eta$ are the gain and the quantum efficiency, respectively, we estimate $G\times\eta$ in excess of $1000$ for these devices.

The results in Fig. \ref{fig:perf}(a)-(b) clearly indicate that the devices are more sensitive at lower incident power density. The strong suppression of responsivity with increasing optical power density is a feature commonly observed in TMDC based photodetectors \cite{Furchi2014}. We are able to fit the photocurrent data using a power law $(I_{ph}\propto P_{op}^\gamma)$, where $\gamma$ is in the range of $0.1$-$0.2$ for different devices. This value is smaller than typical numbers reported for planar TMDC photodetectors \cite{Furchi2014, Island2015}. As mentioned earlier, the gain mechanism in the device is due to (i) confinement of the photo-generated holes the quantum well produced by the heterojunction and (ii) trapping of the photo-generated holes in the band-tail states inside the bandgap of WSe$_2$. At higher intensity of excitation, the trap filling effect gets saturated. In addition, the number density of the confined holes in the valence band of WSe$_2$ quantum well increases leading to enhanced recombination with the photo-generated electrons, further suppressing the gain.

Fig. \ref{fig:perf}(c) shows responsivity numbers for the two types of devices at $V_{ext}=0$. In particular, $D$1 achieves a responsivity of $16.45$ A/W at the lowest power density used. The corresponding specific detectivity ($D^*$) at zero bias is shown in Fig. \ref{fig:perf}(d). $D^*$ is a measure of the ability to sense weak signal. Considering only shot noise from the dark current as the predominant source of noise in the device, specific detectivity for the vertical device can be expressed as:
\begin{equation}\label{eq:Dstar}
D^*=\frac{R}{\sqrt{2qJ_{dark}}}
\end{equation}
where $q$ is the absolute value of electron charge and $J_{dark}$ is the dark current density. Under $V_{ext}=0$, owing to suppression of dark current, we achieve a large specific detectivity ($10^{13}$ Jones for $D$1 and $4.5\times10^{11}$ Jones for $D$2).

\begin{figure*}[!hbt]
\centering
\includegraphics[scale=0.35] {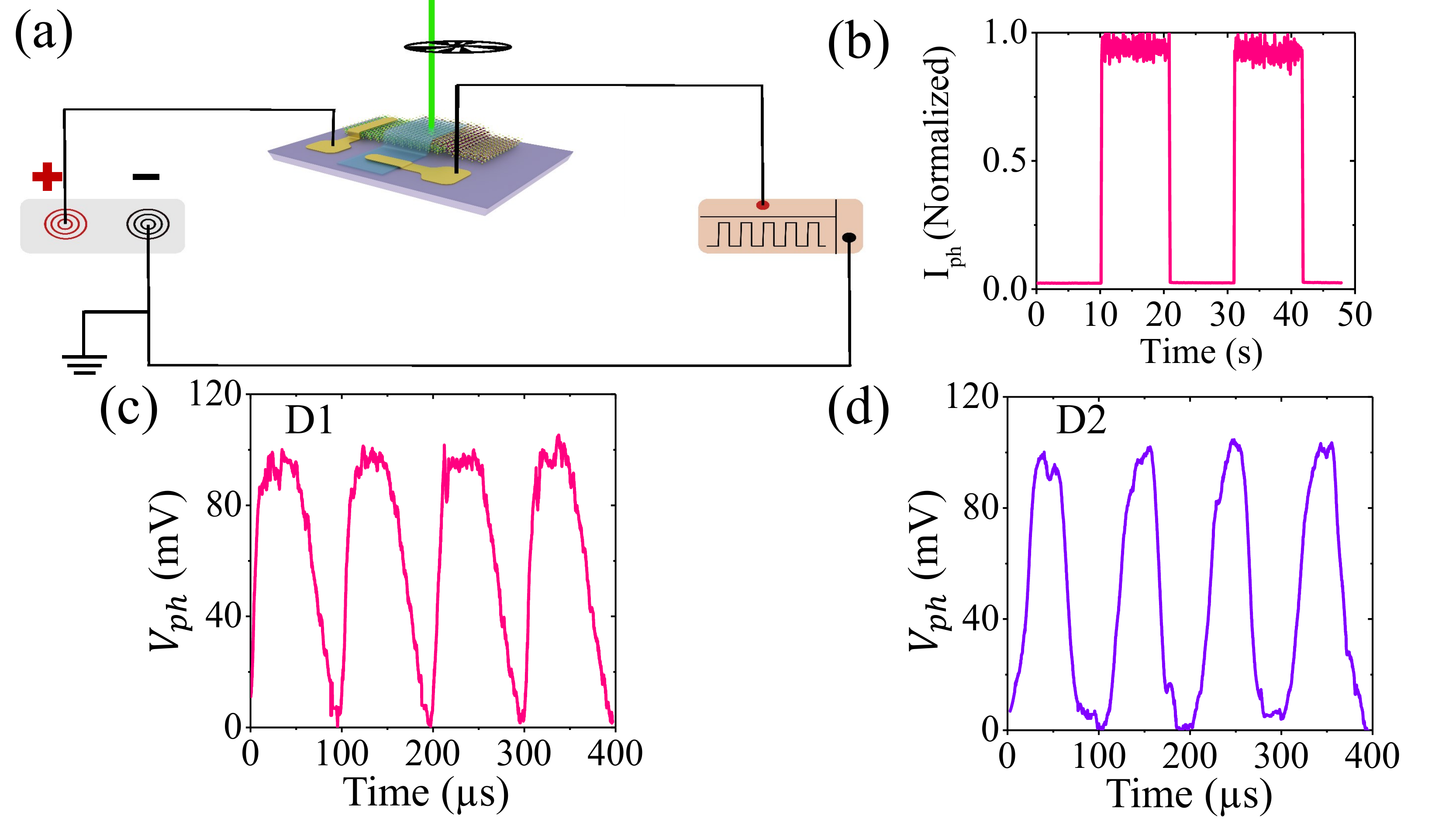}
\caption{(a) Schematic diagram of transient response measurement setup. (b) Transient response of the device $D$1 at relatively longer time scale of seconds showing negligible persistent photocurrent. (c), (d) Temporal response of the photovoltage obtained from $D$1 and $D$2 with $|V_{ext}|=0.2$ V with the mechanical chopper set at its highest frequency of $10$ KHz.}\label{fig:time}
\end{figure*}
We next discuss the transient photoresponse of the devices. A schematic diagram of the experimental setup is illustrated in Fig. \ref{fig:time}(a). A mechanical chopper with $10$ KHz maximum frequency is used to create optical pulses of $50\%$ duty cycle from the continuous wave lasers. A dc voltage source is used to drive the device. The output is measured in a digital storage oscilloscope terminated with $10$ M$\Omega$ impedance. As discussed earlier, planar 2D photodetectors exhibit residual photocurrent when the light source is turned off, due to long lived traps from the interface of SiO$_2$ substrate. In the present vertical structure, such persistent photocurrent is completely avoided, as shown in Fig. \ref{fig:time}(b) in a relatively longer time scale of seconds. In addition, we also observe that the usually observed slow build-up of photocurrent, again due to slow trapping effect, is eliminated. Fig. \ref{fig:time}(c) and (d) depict the transient response of the detectors $D$1 and $D$2, respectively, when the chopper is kept at the maximum frequency of $10$ KHz. The measured \emph{$10\%$-to-$90\%$} rise time ($\tau_r$) and \emph{$90\%$-to-$10\%$} fall time ($\tau_f$), for $D$1 are $8$ $\mu$s and $32$ $\mu$s, respectively. The numbers for $D$2 are $10$ $\mu$s and $13$ $\mu$s, respectively. The results were repeatable over many cycles without any observable ambience induced deterioration.

\begin{figure}[h]
\centering
\includegraphics[scale=0.3] {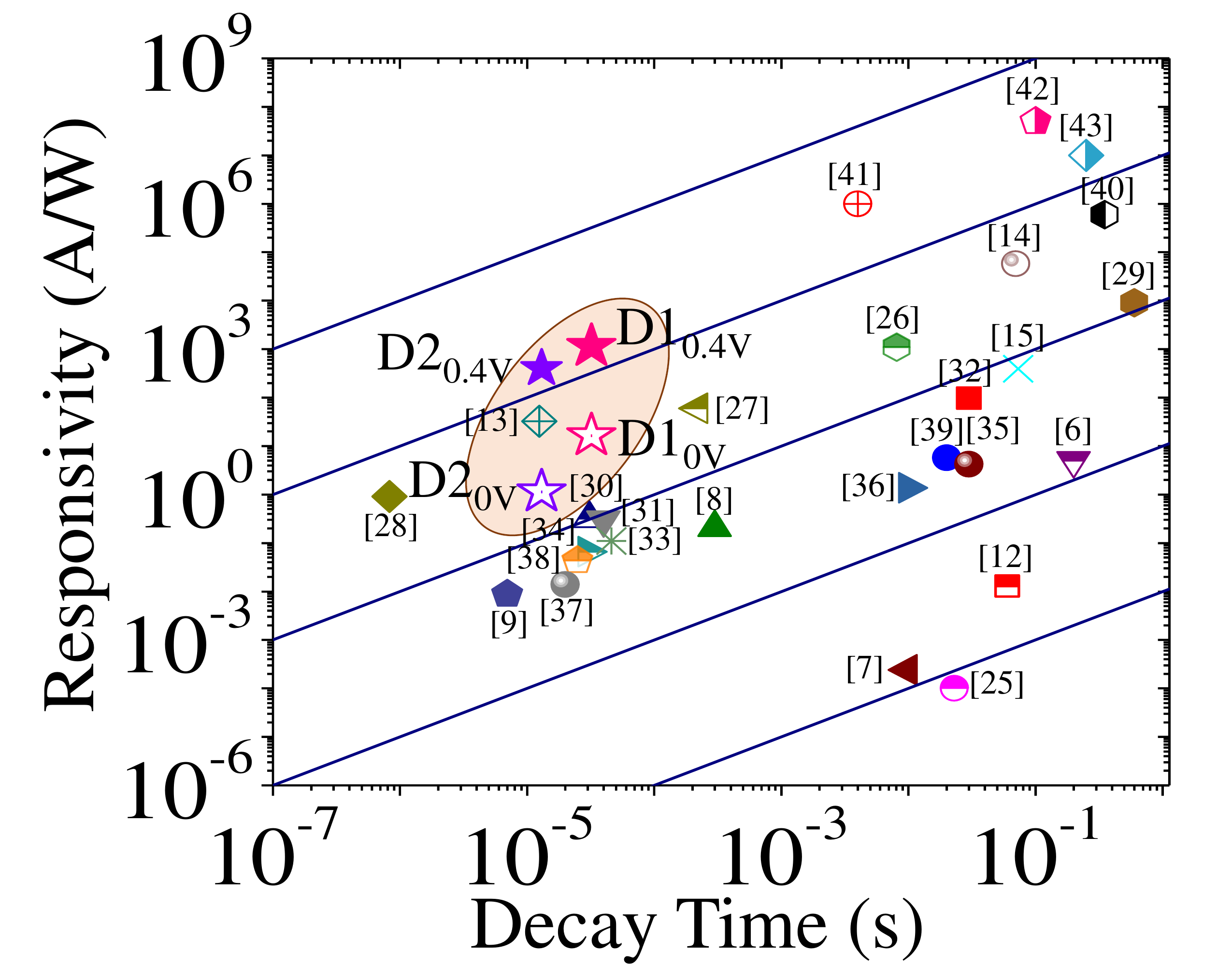}
\vspace{-0.1in}
\caption{Benchmarking the performance of photodetector in responsivity versus $90\%$-to-$10\%$ decay time space. The solid parallel lines correspond to $\frac{\tau_{tr}}{\eta}=63$ ps, $63$ ns, $63$ $\mu$s, $63$ ms, and $63$ s from top to bottom at $\lambda=785$ nm. The data from device $D$1 and $D$2 are represented by open stars ($V_{ext}=0$ V and solid stars ($|V_{ext}|=0.4$ V).}\label{fig:benchmark}
\end{figure}

Finally, we benchmark the performance of the devices presented with other 2-D material based photodetectors from literature in a  responsivity versus fall time chart. The fall time $\tau_f$ after the light source is turned off provides an estimate about the duration the photogenerated hole is trapped in the WSe$_2$ quantum well. Thus the gain can be estimated as $G=\frac{\tau_f}{\tau_{tr}}$, which coupled with (\ref{eq:G}) gives rise to the following relation:
\begin{equation}\label{eq:benchmark_R}
R=\frac{\eta\lambda(nm)}{1243\times\tau_{tr}}\tau_f
\end{equation}
The parallel lines in Fig. \ref{fig:benchmark} are plotted from (\ref{eq:benchmark_R}) for $\lambda=785$ nm with different $\frac{\tau_{tr}}{\eta}$. We populate different points in $R$ versus $\tau_f$ chart from literature \cite{Choi2014, Groenendijk2014, Jacobs-Gedrim2014a, Su2015a, Ghosh2018a, Zhang2014, Zhou2015, Esmaeili-Rad2013, Kallatt2018, Luo2015, Xia2014a, Qiao2018, Zhou2017b, Xu2016, Wang2015, Abderrahmane2014, Zhang2017, Gao2016, Hu2013, Yang2016, Feng2018, Chen2018, Huo2014, Kufer2015, Huo2017, Konstantatos2012, Sun2012}. As expected, the devices with large $R$ also tend to respond slowly. A reduced transit time and improved quantum efficiency can help to improve the performance of the device in the benchmarking chart. The devices reported in this work are represented by open ($V_{ext}=0$ V) and solid ($|V_{ext}|=0.4$ V) stars.  Clearly, these devices exhibit an improved responsivity at a given fall time compared with reports in existing literature.
\section{Conclusion}\label{sec:conclusion}
In conclusion, using systematic optical and electrical probes, we demonstrate efficient control of the charge transfer efficiency and its direction in a skewed and highly staggered ITO/WSe$_2$/SnSe$_2$ double heterojunction. This leads to achievement of high photoconductive gain and high speed photodetection - a unique feature achieved by reduction of transit time of photo-excited electrons using the vertical heterojunction, while confining the photo-excited holes in the WSe$_2$ quantum well. In addition, the built-in electric field in the proposed detector allows photodetection at zero external bias - providing a self-powering feature to the device. The device also exhibits a large open circuit voltage which is promising for solar energy harvesting applications. The device will open a pathway for a novel type of inexpensive, high performance photodetecting devices using vertical heterostructures based on two dimensional materials.
\bibliographystyle{ieeetr}
\bibliography{SWITO}

\begin{thebibliography}{10}

\bibitem{Butler2013}
S.~Z. Butler, S.~M. Hollen, L.~Cao, Y.~Cui, J.~A. Gupta, H.~R. Gutierrez, T.~F.
  Heinz, S.~S. Hong, J.~Huang, A.~F. Ismach, E.~Johnston-Halperin, M.~Kuno,
  V.~V. Plashnitsa, R.~D. Robinson, R.~S. Ruoff, S.~Salahuddin, J.~Shan,
  L.~Shi, M.~G. Spencer, M.~Terrones, W.~Windl, and J.~E. Goldberger,
  ``{Progress, challenges, and opportunities in two-dimensional materials
  beyond graphene},'' {\em ACS Nano}, vol.~7, pp.~2898--2926, mar 2013.

\bibitem{Novoselov2012}
K.~S. Novoselov and A.~H. {Castro Neto}, ``{Two-dimensional crystals-based
  heterostructures: Materials with tailored properties},'' {\em Physica
  Scripta}, vol.~2012, p.~014006, jan 2012.

\bibitem{Geim2013}
A.~K. Geim and I.~V. Grigorieva, ``{Van der Waals heterostructures},'' {\em
  Nature}, vol.~499, pp.~419--425, jul 2013.

\bibitem{Li2016}
M.~Y. Li, C.~H. Chen, Y.~Shi, and L.~J. Li, ``{Heterostructures based on
  two-dimensional layered materials and their potential applications},'' {\em
  Materials Today}, vol.~19, pp.~322--335, aug 2016.

\bibitem{Koppens2014}
F.~H.~L. Koppens, T.~Mueller, P.~Avouris, A.~C. Ferrari, M.~S. Vitiello, and
  M.~Polini, ``{Photodetectors based on graphene, other two-dimensional
  materials and hybrid systems},'' {\em Nature Nanotechnology}, vol.~9,
  pp.~780--793, oct 2014.

\bibitem{Choi2014}
M.~S. Choi, D.~Qu, D.~Lee, X.~Liu, K.~Watanabe, T.~Taniguchi, and W.~J. Yoo,
  ``{Lateral $MoS_2$ p-n junction formed by chemical doping for use in
  high-performance optoelectronics},'' {\em ACS Nano}, vol.~8, pp.~9332--9340,
  aug 2014.

\bibitem{Groenendijk2014}
D.~J. Groenendijk, M.~Buscema, G.~A. Steele, S.~{Michaelis De Vasconcellos},
  R.~Bratschitsch, H.~S. {Van Der Zant}, and A.~Castellanos-Gomez,
  ``{Photovoltaic and photothermoelectric effect in a double-gated $WSe_2$
  device},'' {\em Nano Letters}, vol.~14, pp.~5846--5852, sep 2014.

\bibitem{Esmaeili-Rad2013}
M.~R. Esmaeili-Rad and S.~Salahuddin, ``{High performance molybdenum disulfide
  amorphous silicon heterojunction photodetector},'' {\em Scientific Reports},
  vol.~3, pp.~1--6, aug 2013.

\bibitem{Su2015a}
G.~Su, V.~G. Hadjiev, P.~E. Loya, J.~Zhang, S.~Lei, S.~Maharjan, P.~Dong,
  P.~{M. Ajayan}, J.~Lou, and H.~Peng, ``{Chemical vapor deposition of thin
  crystals of layered semiconductor $SnS_2$ for fast photodetection
  application},'' {\em Nano Letters}, vol.~15, pp.~506--513, dec 2015.

\bibitem{Zheng2016}
L.~Zheng, L.~Zhongzhu, and S.~Guozhen, ``{Photodetectors based on two
  dimensional materials},'' {\em Journal of Semiconductors}, vol.~37,
  p.~091001, sep 2016.

\bibitem{Dhanabalan2016}
S.~C. Dhanabalan, J.~S. Ponraj, H.~Zhang, and Q.~Bao, ``{Present perspectives
  of broadband photodetectors based on nanobelts, nanoribbons, nanosheets and
  the emerging 2D materials},'' {\em Nanoscale}, vol.~8, pp.~6410--6434, feb
  2016.

\bibitem{Xia2014a}
J.~Xia, X.~Huang, L.~Z. Liu, M.~Wang, L.~Wang, B.~Huang, D.~D. Zhu, J.~J. Li,
  C.~Z. Gu, and X.~M. Meng, ``{CVD synthesis of large-area, highly crystalline
  $MoSe_2$ atomic layers on diverse substrates and application to
  photodetectors},'' {\em Nanoscale}, vol.~6, pp.~8949--8955, jun 2014.

\bibitem{Ghosh2018a}
S.~Ghosh, P.~D. Patil, M.~Wasala, S.~Lei, A.~Nolander, P.~Sivakumar, R.~Vajtai,
  P.~Ajayan, and S.~Talapatra, ``{Fast photoresponse and high detectivity in
  copper indium selenide ($CuIn_7Se_{11}$) phototransistors},'' {\em 2D
  Materials}, vol.~5, p.~015001, oct 2018.

\bibitem{Kallatt2018}
S.~Kallatt, S.~Nair, and K.~Majumdar, ``{Asymmetrically Encapsulated Vertical
  $ITO/MoS_2/Cu_2O$ Photodetector with Ultrahigh Sensitivity},'' {\em Small},
  vol.~14, no.~3, p.~1702066, 2018.

\bibitem{Jacobs-Gedrim2014a}
R.~B. Jacobs-Gedrim, M.~Shanmugam, N.~Jain, C.~A. Durcan, M.~T. Murphy, T.~M.
  Murray, R.~J. Matyi, R.~L. Moore, and B.~Yu, ``{Extraordinary photoresponse
  in two-dimensional $In_2Se_3$ nanosheets},'' {\em ACS Nano}, vol.~8,
  pp.~514--521, dec 2014.

\bibitem{Kallatt2016}
S.~Kallatt, G.~Umesh, N.~Bhat, and K.~Majumdar, ``{Photoresponse of atomically
  thin $MoS_2$ layers and their planar heterojunctions},'' {\em Nanoscale},
  vol.~8, no.~33, pp.~15213--15222, 2016.

\bibitem{Krishna2018}
M.~Krishna, S.~Kallatt, and K.~Majumdar, ``{Substrate effects in high gain ,
  low operating voltage $SnSe_2$ photoconductor},'' {\em Nanotechnology},
  vol.~29, p.~035205, dec 2018.

\bibitem{Furchi2014}
M.~M. Furchi, D.~K. Polyushkin, A.~Pospischil, and T.~Mueller, ``{Mechanisms of
  photoconductivity in atomically thin $MoS_2$},'' {\em Nano Letters}, vol.~14,
  pp.~6165--6170, oct 2014.

\bibitem{Murali2018}
K.~Murali, M.~Dandu, S.~Das, and K.~Majumdar, ``{Gate-Tunable $WSe_2 /SnSe_2$
  Backward Diode with Ultrahigh-Reverse Rectification Ratio},'' {\em ACS
  Applied Materials {\&} Interfaces}, vol.~10, pp.~5657--5664, jan 2018.

\bibitem{Somvanshi2017a}
D.~Somvanshi, S.~Kallatt, C.~Venkatesh, S.~Nair, G.~Gupta, J.~K. Anthony,
  D.~Karmakar, and K.~Majumdar, ``{Nature of carrier injection in
  metal/2D-semiconductor interface and its implications for the limits of
  contact resistance},'' {\em Physical Review B}, vol.~96, p.~205423, nov 2017.

\bibitem{Schlaf1999}
R.~Schlaf, O.~Lang, C.~Pettenkofer, and W.~Jaegermann, ``{Band lineup of
  layered semiconductor heterointerfaces prepared by van der Waals epitaxy:
  Charge transfer correction term for the electron affinity rule},'' {\em
  Journal of Applied Physics}, vol.~85, p.~2732, mar 1999.

\bibitem{Lang1994}
O.~Lang, Y.~Tomm, R.~Schlaf, C.~Pettenkofer, and W.~Jaegermann, ``{Single
  crystalline GaSe/$WSe_2$ heterointerfaces grown by van der Waals epitaxy. II.
  Junction characterization},'' {\em Journal of Applied Physics}, vol.~75,
  pp.~7814--7820, feb 1994.

\bibitem{Aretouli2016}
K.~E. Aretouli, D.~Tsoutsou, P.~Tsipas, J.~Marquez-Velasco, S.~{Aminalragia
  Giamini}, N.~Kelaidis, V.~Psycharis, and A.~Dimoulas, ``{Epitaxial 2D
  $SnSe_2$/ 2D $WSe_2$ van der Waals heterostructures},'' {\em ACS Applied
  Materials and Interfaces}, vol.~8, pp.~23222--23229, aug 2016.

\bibitem{Island2015}
J.~O. Island, S.~I. Blanter, M.~Buscema, H.~S. {Van Der Zant}, and
  A.~Castellanos-Gomez, ``{Gate controlled photocurrent generation mechanisms
  in high-gain $In_2Se_3$ phototransistors},'' {\em Nano Letters}, vol.~15,
  pp.~7853--7858, nov 2015.

\bibitem{Zhang2014}
W.~Zhang, M.~H. Chiu, C.~H. Chen, W.~Chen, L.~J. Li, and A.~T.~S. Wee, ``{Role
  of metal contacts in high-performance phototransistors based on $WSe_2$
  monolayers},'' {\em ACS Nano}, vol.~8, pp.~8653--8661, aug 2014.

\bibitem{Zhou2015}
X.~Zhou, L.~Gan, W.~Tian, Q.~Zhang, S.~Jin, H.~Li, Y.~Bando, D.~Golberg, and
  T.~Zhai, ``{Ultrathin $SnSe_2$ Flakes Grown by Chemical Vapor Deposition for
  High-Performance Photodetectors},'' {\em Advanced Materials}, vol.~27,
  pp.~8035--8041, nov 2015.

\bibitem{Luo2015}
W.~Luo, Y.~Cao, P.~Hu, K.~Cai, Q.~Feng, F.~Yan, T.~Yan, X.~Zhang, and K.~Wang,
  ``{Gate Tuning of High-Performance InSe-Based Photodetectors Using Graphene
  Electrodes},'' {\em Advanced Optical Materials}, vol.~3, pp.~1418--1423, jun
  2015.

\bibitem{Qiao2018}
S.~Qiao, R.~Cong, J.~Liu, B.~Liang, G.~Fu, W.~Yu, K.~Ren, S.~Wang, and C.~Pan,
  ``{Vertical layered $MoS_2/Si$ heterojunction for ultrahigh and ultrafast
  photoresponse photodetector},'' {\em Journal of Materials Chemistry C}, p.~1,
  feb 2018.

\bibitem{Zhou2017b}
X.~Zhou, N.~Zhou, C.~Li, H.~Song, Q.~Zhang, X.~Hu, L.~Gan, H.~Li, J.~L{\"{u}},
  J.~Luo, J.~Xiong, and T.~Zhai, ``{Vertical heterostructures based on $SnSe_2
  /MoS_2$ for high performance photodetectors},'' {\em 2D Materials}, vol.~4,
  p.~025048, mar 2017.

\bibitem{Xu2016}
Z.~Xu, S.~Lin, X.~Li, S.~Zhang, Z.~Wu, W.~Xu, Y.~Lu, and S.~Xu, ``{Monolayer
  $MoS_2/GaAs$ heterostructure self-driven photodetector with extremely high
  detectivity},'' {\em Nano Energy}, vol.~23, pp.~89--96, mar 2016.

\bibitem{Wang2015}
L.~Wang, J.~Jie, Z.~Shao, Q.~Zhang, X.~Zhang, Y.~Wang, Z.~Sun, and S.~T. Lee,
  ``{$MoS_2/Si$ heterojunction with vertically standing layered structure for
  ultrafast, high-detectivity, self-driven visible-near infrared
  photodetectors},'' {\em Advanced Functional Materials}, vol.~25,
  pp.~2910--2919, mar 2015.

\bibitem{Abderrahmane2014}
A.~Abderrahmane, P.~J. Ko, T.~V. Thu, S.~Ishizawa, T.~Takamura, and A.~Sandhu,
  ``{High photosensitivity few-layered $MoSe_2$ back-gated field-effect
  phototransistors},'' {\em Nanotechnology}, vol.~25, p.~365202, aug 2014.

\bibitem{Zhang2017}
K.~Zhang, X.~Fang, Y.~Wang, Y.~Wan, Q.~Song, W.~Zhai, Y.~Li, G.~Ran, Y.~Ye, and
  L.~Dai, ``{Ultrasensitive Near-Infrared Photodetectors Based on a
  Graphene-$MoTe_2$-Graphene Vertical van der Waals Heterostructure},'' {\em
  ACS Applied Materials and Interfaces}, vol.~9, pp.~5392--5398, jan 2017.

\bibitem{Gao2016}
A.~Gao, E.~Liu, M.~Long, W.~Zhou, Y.~Wang, T.~Xia, W.~Hu, B.~Wang, and F.~Miao,
  ``{Gate-tunable rectification inversion and photovoltaic detection in
  $graphene/WSe_2$ heterostructures},'' {\em Applied Physics Letters},
  vol.~108, p.~223501, jun 2016.

\bibitem{Hu2013}
P.~Hu, L.~Wang, M.~Yoon, J.~Zhang, W.~Feng, X.~Wang, Z.~Wen, J.~C. Idrobo,
  Y.~Miyamoto, D.~B. Geohegan, and K.~Xiao, ``{Highly responsive ultrathin GaS
  nanosheet photodetectors on rigid and flexible substrates},'' {\em Nano
  Letters}, vol.~13, pp.~1649--1654, mar 2013.

\bibitem{Yang2016}
S.~Yang, C.~Wang, C.~Ataca, Y.~Li, H.~Chen, H.~Cai, A.~Suslu, J.~C. Grossman,
  C.~Jiang, Q.~Liu, and S.~Tongay, ``{Self-Driven Photodetector and Ambipolar
  Transistor in Atomically Thin GaTe-$MoS_2$ p-n vdW Heterostructure},'' {\em
  ACS Applied Materials and Interfaces}, vol.~8, no.~4, pp.~2533--2539, 2016.

\bibitem{Feng2018}
W.~Feng, Z.~Jin, J.~Yuan, J.~Zhang, S.~Jia, L.~Dong, J.~Yoon, L.~Zhou,
  R.~Vajtai, J.~M. Tour, and P.~M. Ajayan, ``{A fast and zero-biased
  photodetector based on GaTe – InSe vertical 2D p – n heterojunction},''
  {\em 2D Materials}, vol.~5, p.~025008, feb 2018.

\bibitem{Chen2018}
Y.~Chen, X.~Wang, G.~Wu, Z.~Wang, H.~Fang, T.~Lin, S.~Sun, H.~Shen, W.~Hu,
  J.~Wang, J.~Sun, X.~Meng, and J.~Chu, ``{High-Performance Photovoltaic
  Detector Based on $MoTe_2/MoS_2$ Van der Waals Heterostructure},'' {\em
  Small}, vol.~14, p.~1703293, jan 2018.

\bibitem{Huo2014}
N.~Huo, S.~Yang, Z.~Wei, S.-S. Li, J.-B. Xia, and J.~Li, ``{Photoresponsive and
  Gas Sensing Field-Effect Transistors based on Multilayer $WS_2$
  Nanoflakes},'' {\em Scientific Reports}, vol.~4, p.~5209, jun 2014.

\bibitem{Kufer2015}
D.~Kufer, I.~Nikitskiy, T.~Lasanta, G.~Navickaite, F.~H. Koppens, and
  G.~Konstantatos, ``{Hybrid 2D-0D $MoS_2$-PbS quantum dot photodetectors},''
  {\em Advanced Materials}, vol.~27, pp.~176--180, nov 2015.

\bibitem{Huo2017}
N.~Huo, S.~Gupta, and G.~Konstantatos, ``{Mo$S_2$–HgTe Quantum Dot Hybrid
  Photodetectors beyond 2 µm},'' {\em Advanced Materials}, vol.~29,
  p.~1606576, mar 2017.

\bibitem{Konstantatos2012}
G.~Konstantatos, M.~Badioli, L.~Gaudreau, J.~Osmond, M.~Bernechea, F.~P.~G. {De
  Arquer}, F.~Gatti, and F.~H. Koppens, ``{Hybrid graphene-quantum dot
  phototransistors with ultrahigh gain},'' {\em Nature Nanotechnology}, vol.~7,
  pp.~363--368, may 2012.

\bibitem{Sun2012}
Z.~Sun, Z.~Liu, J.~Li, G.~A. Tai, S.~P. Lau, and F.~Yan, ``{Infrared
  photodetectors based on CVD-grown graphene and PbS quantum dots with
  ultrahigh responsivity},'' {\em Advanced Materials}, vol.~24, pp.~5878--5883,
  aug 2012.

\end{thebibliography}
\end{document}